\pdfoutput=1

\documentclass{article}

\PassOptionsToPackage{authoryear,round}{natbib}
\usepackage[preprint]{neurips_2025}

\usepackage[utf8]{inputenc}
\usepackage[T1]{fontenc}
\usepackage{hyperref}
\usepackage{url}
\usepackage{booktabs}
\usepackage{amsfonts}
\usepackage{amsmath}
\usepackage{nicefrac}
\usepackage{microtype}
\usepackage{xcolor}
\definecolor{OliveGreen}{rgb}{0.20, 0.55, 0.20}
\usepackage{graphicx}
\usepackage{colortbl}
\usepackage{enumitem}
\usepackage{subcaption}
\usepackage{multirow}
\usepackage{array}
\usepackage{float}
\usepackage{makecell}
\usepackage{xspace}

\newcommand{\sysname}{\textsc{LuminaECG}\xspace}
\newtcolorbox{finding}[1][]{
  colback=black!5!white,    
  colframe=black!60!white,  
  fonttitle=\bfseries,     
  title=Finding,          
  arc=4pt,                 
  outer arc=4pt,
  #1                       
}

\hypersetup{
    colorlinks=true,
    citecolor=neuripsaccentblue,
    linkcolor=neuripsaccentblue,
    urlcolor=neuripsaccentblue
}

\title{Diagnosing as Cardiologists Do: ECG Agents with Doctor-Grounded Priors for Clinical Reasoning Across Diseases and Populations}

\author{%
  \textbf{Hongxiang Gao\textsuperscript{1,2,*}},
  \textbf{He-Yang Xu\textsuperscript{1,2,*}},
  \textbf{Yuwen Li\textsuperscript{1,2}},
  \textbf{Minghui Zhao\textsuperscript{1,2}},
  \textbf{Zhipeng Cai\textsuperscript{1,2}},\\
  \textbf{Xingyao Wang\textsuperscript{1,2}},
  \textbf{Chenxi Yang\textsuperscript{1,2}},
  \textbf{Jianqing Li\textsuperscript{1,2}},
  \textbf{Chengyu Liu\textsuperscript{1,2,\#,\textdagger}}\\
  \small $^1$ State Key Laboratory of Digital Medical Engineering, Southeast University \\
  \small $^2$ School of Instrument Science and Engineering, Southeast University \\
  \small $^*$ Equal Core Contributions, $^\#$ Project Lead, $^\dagger$ Corresponding Author \\
  \small \texttt{\{hongxiang\_seu, xuhy, chengyu\}@seu.edu.cn}
}

\begin{document}
\maketitle

\begin{abstract}

Cardiologists interpret electrocardiograms by localizing waveform components, measuring rhythm and interval patterns, and translating these structured observations into diagnostic evidence. Whether this expert reading process can serve as an effective prior for ECG agents remains unclear. To address this question, we introduce \sysname, a clinically structured ECG reasoning framework that reformulates ECG interpretation as measurement-grounded visual reading. ECG signals are rendered on standard electrocardiographic grid paper to preserve the spatial and scale cues used in clinical reading. P-wave, QRS-complex, and T-wave boundaries are explicitly delineated, and color-coded segmentation decomposes the waveform into discrete visual measurement primitives. A general 2B vision–language backbone is then trained with low-rank supervised fine-tuning to associate these primitives with diagnostic reasoning, without architectural modification. Across open, proprietary, and ECG-specialist zero-shot baselines, \sysname improves both waveform measurement and diagnostic recovery. It reaches a clinically meaningful reader tier on the CODE-test benchmark, transfers across geographically diverse ECG datasets without retraining, and generates reports whose structure contains an emergent prognostic signal. These findings suggest that effective ECG agents require not only larger models, but supervision that preserves the alignment between measurable waveform evidence and clinical knowledge.
\\
\\
\medskip
\begin{flushleft}
  \begin{tabular}{@{}ll@{}}
     \includegraphics[width=1em]{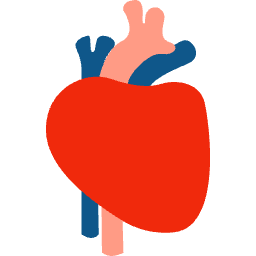} \textbf{Project Page} & \url{https://luminaecg.github.io} \\
    \includegraphics[width=1em]{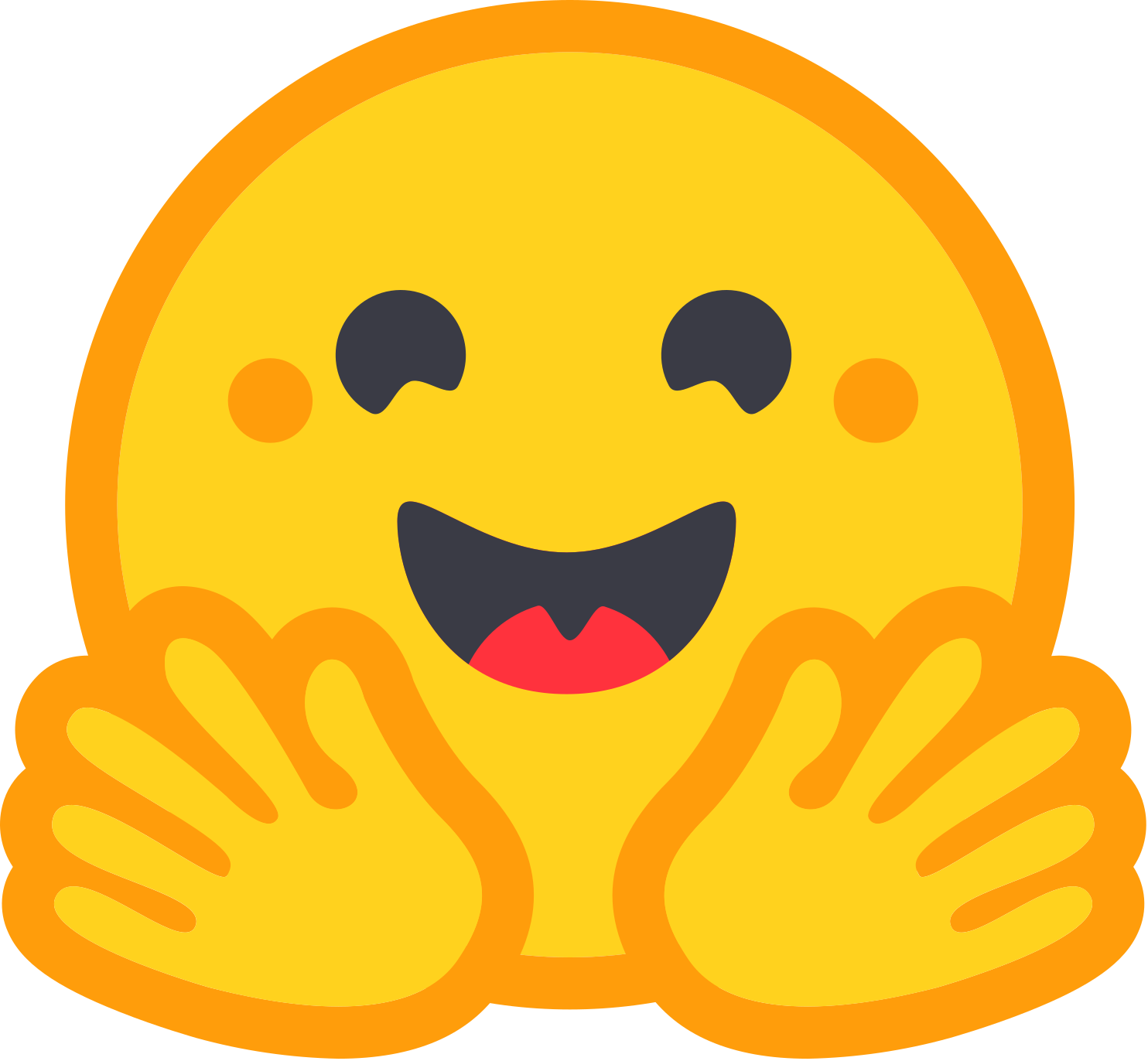} \textbf{Model} & \url{https://huggingface.co/hiangx/LuminaECG} \\
  \end{tabular}
\end{flushleft}
\end{abstract}

\clearpage
\section{Introduction}
\label{sec:intro}
\begin{figure}[!tbp]
    \centering
    \includegraphics[width=1.0\linewidth]{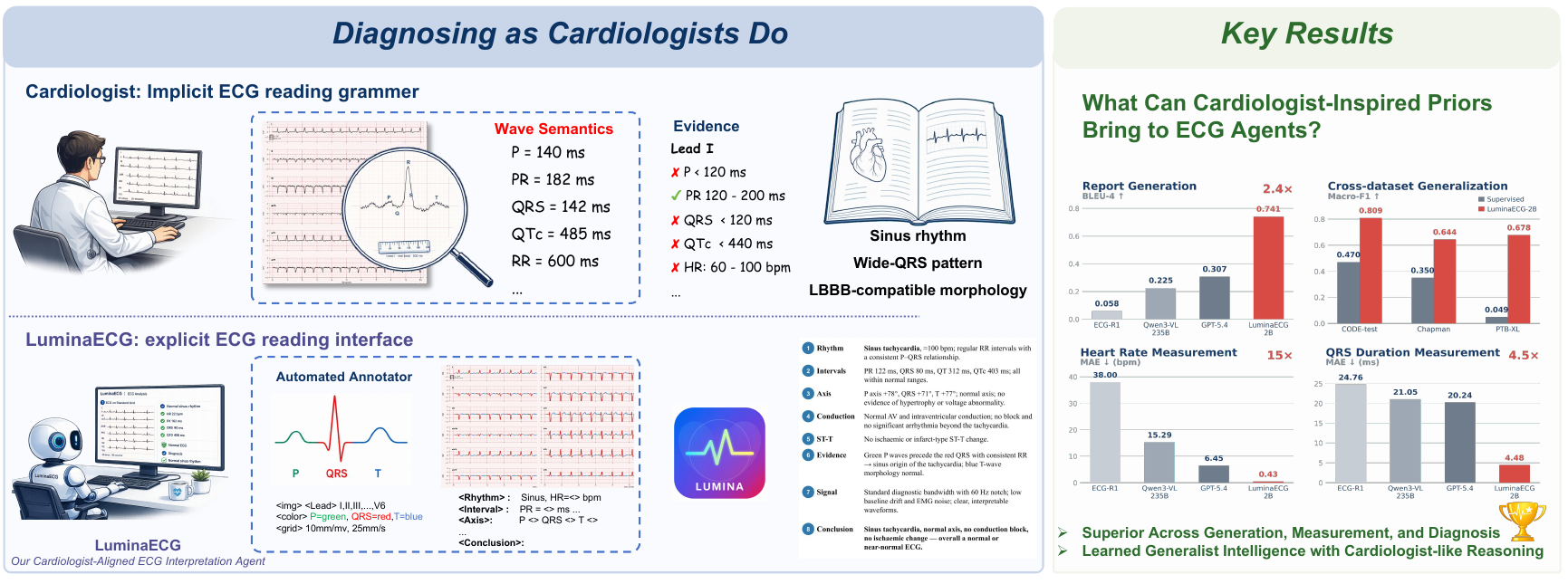}
    \caption{\footnotesize Overview of LuminaECG. Cardiologist diagnostic priors are encoded into the data construction process to guide waveform localization, measurement, and evidence-based interpretation, yielding consistent gains in report generation, quantitative accuracy, and cross-dataset generalization.}
    \label{fig:teaser}
\end{figure}

The electrocardiogram~(ECG) is one of the most information-dense measurements in routine clinical practice. 
 A ten-second, twelve-lead ECG encodes several distinct kinds of cardiac information at once: rhythm, conduction intervals, electrical axis, chamber activity, and lead-specific ST–T morphology. Experts reach a diagnosis through a systematic measurement of each waveform component against the calibrated grid, where every small square stands for a fixed interval of time and voltage~\citep{aha2009ecg}. Rate, intervals, axis, and ST–T geometry are measured off this grid and matched to diagnostic criteria~\citep{aha2009ischemia,aha2009hypertrophy,aha2009ivcd}, so that each conclusion stays anchored to a quantity that can be checked (cf. Fig.~\ref{fig:teaser}). This measurement-grounded integration of evidence is what turns a printed tracing into a clinical diagnosing, and it sets the standard that any automated reader must meet.

As artificial intelligence has advanced, a growing number of methods have been developed for automated ECG interpretation~\citep{hannun2019,ribeiro2020,attia2019af,attia2019lowef}. Conventional supervised ECG models perform well on predefined tasks but remain limited in generalization beyond fixed label spaces~\citep{ballas2023,strodthoff2021,siontis2021}. Multimodal large language models (MLLMs) provide a more flexible alternative and have shown rapidly improving capabilities across medical diagnostic tasks~\citep{llavamed2023,medpalmm2024}. However, these models lack the domain-specific priors required for detailed ECG interpretation, including waveform localization, interval measurement, and the correspondence between morphological evidence and diagnostic criteria~\citep{meit2024,ecgqa2023,pulse2026,gem2025,ecgr1}. A model may therefore know what atrial fibrillation or bundle-branch block means, yet still fail to identify the evidence supporting that diagnosis in a particular ECG. Despite these advances, current ECG models remain difficult to deploy reliably in real-world clinical settings, where predictions must generalize and remain grounded in waveform evidence. This leads to a broader question: what should the next generation of ECG agents look like?

\emph{Better diagnostic priors from cardiologists are all we need.} Guided by this idea, we introduce \sysname, an ECG reasoning framework designed to test whether the structure of clinical interpretation can elevate diagnostic performance toward the level of human readers. Following the workflow of cardiologists, we first build an agentic data-processing pipeline that renders each twelve-lead ECG on standardized grid paper, preserving the temporal and voltage scales required for clinical measurement (cf. Fig.~\ref{fig:teaser}). The P wave, QRS complex, and T wave are then explicitly delineated~\citep{martinez2004,densecg2021,joung2024}, making key waveform components directly identifiable to the model. Signal-derived measurements and available annotations are further organized into structured reports that mirror the progression from waveform observation and quantitative assessment to diagnostic conclusion. Rather than relying on a large and complex multimodal language model, we deliberately fine-tune a compact 2B backbone~\citep{qwen3vl,lora} without architectural modification or additional alignment procedures. This design allows us to directly test whether clinically informed data construction, rather than model scaling alone, can provide a stronger foundation for ECG interpretation.

We conduct extensive experiments to evaluate whether clinician-informed supervision can compensate for limited model scale. Despite using only a compact 2B backbone and a simple fine-tuning procedure, \sysname consistently outperforms the evaluated open, proprietary, and ECG-specialist zero-shot baselines in waveform measurement, diagnostic recovery, and structured report generation. The improvement is particularly pronounced for quantitative interpretation: \sysname achieves a heart-rate mean absolute error of only 0.43 bpm, approximately fifteen times lower than the best error among the compared zero-shot models. On the CODE-test benchmark~\citep{ribeiro2020}, it attains a macro-F1 of 0.840 and a micro-F1 of 0.852, exceeding both the medical-student tier (0.817 and 0.833) and the emergency-resident tier (0.829 and 0.846), and approaching the cardiology-resident tier (0.862 and 0.876). The model also transfers without retraining across ECG datasets collected from geographically distinct health systems. Beyond diagnostic performance, its generated reports retain prognostic information in an unseen cohort, despite the model not being trained for outcome prediction.

Together, these results suggest that the bottleneck in medical multimodal modelling lies not only in model capacity, but also in whether the training data preserve how clinicians localize evidence, derive measurements, and connect those measurements to diagnostic conclusions. For ECG, and potentially other measurement-intensive medical domains, expert-informed data design may therefore be a more fundamental step than model scaling. Before building larger medical models, we may first need to build data that preserve how clinicians see, measure, and reason.

\section{Related Works}
\label{sec:related}

\paragraph{Deep learning for automatic ECG diagnosis.}
Deep learning has driven the main advances in automatic ECG diagnosis, moving the field from
hand-crafted-feature classifiers~\citep{dechazal2004,martinez2004} to networks that learn from the raw
signal and reach cardiologist-level accuracy on rhythm and multi-label
tasks~\citep{hannun2019,ribeiro2020,siontis2021}, with later architectural and self-supervised advances
broadening their coverage~\citep{densecg2021,joung2024,mina2019,yao2020,stmem2024,ecgfm2024,merl2024,kmerl2025}.
These systems are closed-set classifiers: they emit a label rather than an interpretation, expose no
localized evidence for a call, and lose much of their in-distribution accuracy across recorders,
populations, and acquisition conditions, which keeps them short of dependable unsupervised clinical
use~\citep{ballas2023,strodthoff2021,ptbxl,siontis2021}.

\paragraph{Foundation models and multimodal LLMs for ECG.}
To move past closed-set labels, a second line couples the tracing with large language models, first
pairing a signal encoder with a text decoder for report generation and question
answering~\citep{meit2024,ecgqa2023,ecgchat2024} and, closest to our setting, reading the paper-format
twelve-lead ECG directly as an image~\citep{pulse2026,gem2025} in the manner of general medical
vision-language models~\citep{llavamed2023,medpalmm2024}; ECG-R1 goes further, shaping the output with
diagnostic protocols and reinforcement rewards~\citep{ecgr1}. What these systems supervise is report
text, and the correspondence between a stated finding and the waveform evidence for it is never made
explicit. A report can therefore read fluently and still disagree with the tracing it describes, a
failure quantified for radiology report generation~\citep{r2gen2020,miura2021,radgraph2021}: nothing
ties a diagnostic claim to the waveform component that supports it or the measurement that quantifies
it.

\paragraph{Injecting prior knowledge into learning.}
A broad literature makes expert priors visible to a model, whether in the
loss~\citep{raissi2019,karniadakis2021piml}, the
curriculum~\citep{bengio2009curriculum,kumar2010}, the supervised reasoning
trace~\citep{wei2022cot,lightman2024verify}, the pixels via visual prompts that ring a region of
interest~\citep{shtedritski2023redcircle,som2023,vipllava2024,wan2024crg}, or, for ECG, attention and
knowledge-enhanced objectives~\citep{mina2019,kmerl2025}. Each adds one prior at one level; grounded ECG
reading needs three registered together in a single training example: a metric coordinate frame, the
identity of each waveform component, and the tie from a measurement to the diagnosis it supports.
\sysname builds all three into every training example, rendering the ECG on its native paper grid,
colour-coding the P, QRS, and T waves, and writing a measurement-grounded structured report, and fits
these aligned views with ordinary low-rank fine-tuning~\citep{qwen3vl,lora}. We treat this visible-prior
ladder as a hypothesis, and test below whether the model reads from these priors or from correlated
shortcuts.

\section{Results}
We evaluate LuminaECG with a unified protocol designed to make comparisons across ECG agents clinically meaningful. The protocol fixes the test cohorts, input format, diagnostic endpoints, and reporting criteria, so that performance differences reflect ECG interpretation ability rather than variations in prompting, preprocessing, or label construction.

The evaluation is organized around three capabilities required for clinically ECG diagnose and interpretation. We first test whether LuminaECG can generate doctor-style reports with higher linguistic fidelity than general vision--language models. We then ask whether these reports are grounded in quantitative ECG measurements, including heart rate, PR interval, and QRS duration. Finally, we examine whether doctor-grounded supervision supports clinically meaningful interpretation, both by comparison against human readers of different expertise tiers on CODE-test and by transferring across external cohorts. This diagnostic stance motivates the structure of the following results.

\subsection{Evaluation Protocol}
We evaluated LuminaECG with a unified protocol designed to test whether clinical reading priors can move an ECG agent toward human-level interpretation while preserving interpretable evidence. The protocol covers four complementary aspects: generated report quality, clinically essential numeric measurements, comparison with human readers on CODE-test, and cross-cohort transfer after doctor-style supervision.

For report generation and numeric measurement, all models were tested on the same held-out MIMIC-IV-ECG set of $79{,}981$ records, using the same ECG-derived inputs, guideline-structured reference reports, and waveform-derived numeric references. Report quality \cite{bleu2002,cider2015} was measured with BLEU, ROUGE-L and CIDEr to quantify agreement with clinical references. Measurement fidelity was measured by mean absolute error for heart rate, PR interval and QRS duration. All values are averaged over the full test set, so differences between methods reflect reporting and measurement ability rather than differences in data, prompting, or evaluation rules.

For diagnostic interpretation, we used macro-F1 and micro-F1 on clinically defined ECG conditions. CODE-test \cite{ribeiro2020} provides a human-reader scale, allowing the agent to be compared with medical students, residents, cardiologists and a published deep-learning baseline. Cross-cohort evaluation further tests whether the learned clinical reading priors transfer to external datasets with no training contact. Macro-F1 gives equal weight to each diagnosis, whereas micro-F1 reflects aggregate performance across all decisions.

\subsection{Doctor-grounded supervision yields measurement-grounded ECG reporting}
For an ECG agent to approach human-like diagnosis, the prediction itself is not sufficient: the model must also produce interpretable evidence that resembles the way clinicians read ECGs. A correct label without a faithful report may reflect shortcut learning, while a fluent report without accurate measurements may only imitate clinical language. We therefore first evaluated whether doctor-grounded supervision enables LuminaECG to generate guideline-structured ECG reports and to preserve the quantitative measurements that support clinical interpretation, including heart rate, PR interval and QRS duration.

On the MIMIC-IV-ECG held-out test set ($n=79{,}981$), LuminaECG showed a large and consistent advantage in report-generation quality. It achieved the best score on every text-similarity metric, with BLEU-1 $0.853$, BLEU-4 $0.741$, ROUGE-L $0.808$ and CIDEr $0.896$. In contrast, the strongest general frontier systems remained substantially lower: GPT-5.4 reached BLEU-4 $0.307$ and ROUGE-L $0.392$, while Claude-Opus-4.7 reached BLEU-4 $0.292$ and ROUGE-L $0.383$. The gap was especially pronounced on CIDEr, where LuminaECG reached $0.896$ whereas all compared zero-shot systems were near zero. This indicates that the improvement is not limited to local lexical overlap, but extends to the report-level structure and clinically specific phrasing induced by the doctor-style supervision.

The same pattern appeared in numeric measurement fidelity. LuminaECG reduced heart-rate MAE to $0.43$ bpm, compared with $6.45$ bpm for GPT-5.4, $8.37$ bpm for Claude-Opus-4.7 and $15.73$ bpm for Qwen3-VL-8B. For ECG intervals, LuminaECG also achieved the lowest error, with PR MAE $8.97$ ms and QRS MAE $4.48$ ms, whereas the strongest zero-shot systems remained at approximately $24$--$26$ ms for PR and $14$--$21$ ms for QRS. Thus, doctor-grounded supervision does more than make the report text resemble clinical writing: it anchors the generated interpretation to waveform-derived measurements that clinicians rely on when forming diagnoses. 

Together, these results show that LuminaECG learns measurement-grounded ECG reporting rather than superficial report imitation. The generated outputs are both more faithful to guideline-structured clinical references and more accurate in the numeric evidence required for ECG interpretation, providing the basis for comparing the agent with human readers and testing whether the learned clinical reading priors transfer across cohorts.

\begin{finding}[title=Finding 1]
Doctor-grounded supervision converts ECG reporting from language imitation into measurement-grounded interpretation.
\end{finding}

\subsection{Frontier systems miss critical findings; doctor-grounded supervision closes the gap}

Beyond report fluency and measurement accuracy, a clinically useful ECG agent must not overlook findings that carry immediate clinical consequence; a model that produces accurate normal reports but misses life-threatening abnormalities is unsafe regardless of its aggregate scores. We therefore evaluated critical-finding recall, measuring how many pre-specified urgent diagnoses each system recovers.

On PTB-XL ($n=4{,}396$, no training contact), across $2{,}569$ critical-finding instances spanning eleven urgent diagnoses (atrial fibrillation, atrial flutter, ventricular tachycardia, third-degree AV block, LBBB, RBBB, ST elevation, ST depression, QTc prolongation, old infarction, and LV hypertrophy), \sysname-2B recovered $50.8\%$, whereas every frontier zero-shot system remained below $10.3\%$: Claude-Opus-4.7 reached $0.096$, GPT-5.4 $0.076$, Gemini-3.1-Pro $0.051$, and the largest open model, Qwen3-VL-235B-A22B, only $0.008$. A 2-billion-parameter doctor-grounded model thus exceeded a 235-billion-parameter zero-shot model on this endpoint by roughly $64\times$. The gap is not closed by ECG specialisation either: on in-domain MIMIC-IV-ECG the published ECG-specialist ECG-R1 recovered only $1.1\%$ of critical findings, missing $81\%$. This indicates that the bottleneck for recovering critical findings is supervision design rather than parameter count or domain pre-training.

The advantage was uniform across conditions rather than driven by a single class: \sysname-2B was the strongest system on eight of the eleven critical findings, including atrial fibrillation ($0.92$ recall versus $0$--$0.16$ for the next-best frontier), LV hypertrophy ($0.78$ versus $0$--$0.28$) and LBBB ($0.70$ versus $0$--$0.06$). The systems also differed in how they failed, with frontier models separating into loud hallucinators (Claude, GPT and the Qwen family, $63$--$72\%$ hallucinated claims) and a silent omitter (Gemini, $64\%$ of findings omitted), whereas \sysname-2B alone occupied a balanced low-error regime (hallucination $0.21$, omission $0.21$). Together, these results show that doctor-grounded supervision closes the critical-finding safety gap separating general vision--language models from clinical ECG interpretation, and that scale alone does not narrow it.

\begin{finding}[title=Finding 2]
Critical-finding recovery depends more on clinically grounded supervision than on model scale. Despite using only a 2B backbone, \sysname recovered $50.8\%$ of urgent findings on unseen PTB-XL data, whereas all evaluated frontier zero-shot systems remained below $10.3\%$ and showed either severe hallucination or omission.
\end{finding}

\subsection{Clinical diagnostic priors bring ECG agents closer to human readers}

A clinically meaningful ECG agent should be evaluated on a scale that is interpretable to clinicians. Instead of reporting diagnostic scores only against other algorithms, we used CODE-test as a human-reader benchmark, where the same six CODE-6 red-flag diagnoses are evaluated across medical students, residents, cardiologists, and the Ribeiro-2020 deep-convolutional baseline. This setting allows us to ask whether clinical diagnostic priors learned from doctor-style supervision can move an ECG agent toward human reader performance.

\begin{figure*}[t]
  \centering
  \begin{subfigure}[t]{0.48\linewidth}
    \centering
    \includegraphics[width=\linewidth]{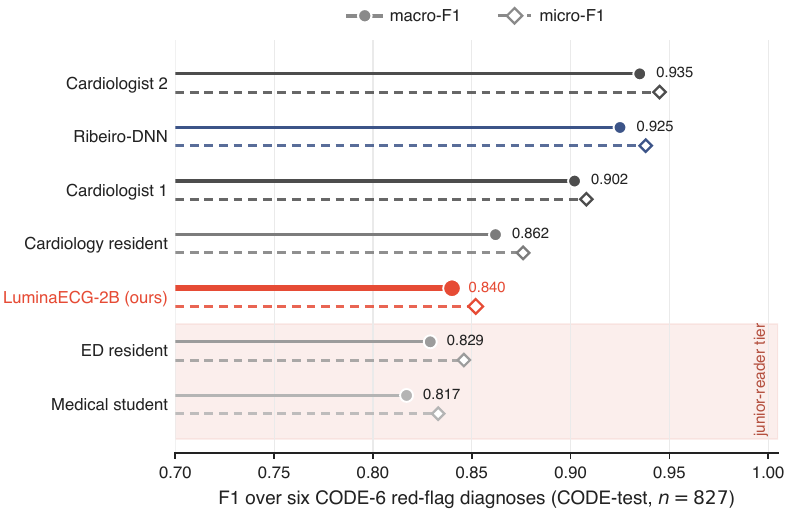}
    \caption{\textbf{Comparison with human readers on CODE-test.}
Macro-F1 (solid) and micro-F1 (dashed) across six CODE-6 red-flag diagnoses. \sysname-2B exceeds the medical-student and emergency-resident tiers, and remains below the cardiology resident, the cardiologists, and the Ribeiro-2020 baseline.}
    \label{fig:exp-vshuman}
  \end{subfigure}
  \hfill
  \begin{subfigure}[t]{0.48\linewidth}
    \centering
    \includegraphics[width=\linewidth]{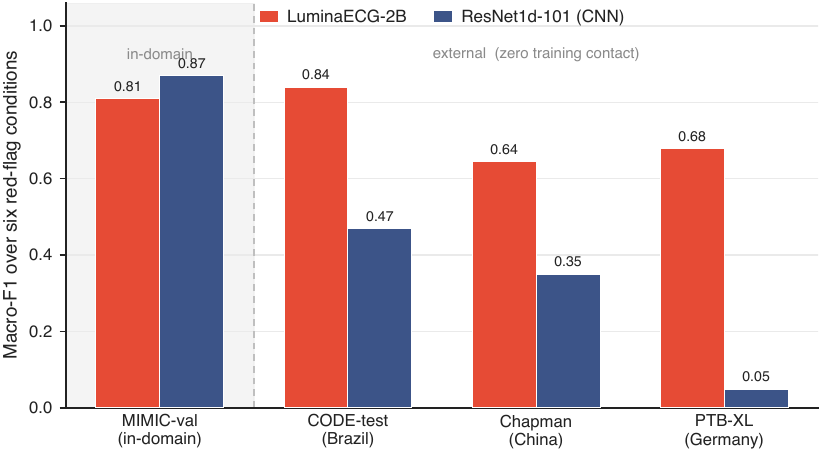}
    \caption{\textbf{Cross-cohort generalization without target-cohort training.}
Macro-F1 across six red-flag diagnoses for \sysname-2B and a ResNet1d-101. Although comparable in-domain, the CNN degrades more sharply across external cohorts, whereas \sysname-2B remains more stable.}
    \label{fig:exp-transport}
  \end{subfigure}

  \caption{\footnotesize \textbf{\sysname-2B surpasses the junior-reader tiers and generalizes across external cohorts.} \textbf{(a)} Performance on the CODE-test reader benchmark. \textbf{(b)} Cross-cohort comparison with a CNN trained on the same source data. Full results are provided in Tables~\ref{tab:app-vshuman} and~\ref{tab:app-crossdata}.}
  \label{fig:exp-human-transport}
\end{figure*}

On CODE-test ($n=827$), \sysname-2B reached macro-F1 $0.840$ and micro-F1 $0.852$ over the six red-flag diagnoses (cf. Fig.~\ref{fig:exp-vshuman}). This places the model above the medical-student tier (macro-F1 $0.817$) and the emergency-resident tier ($0.829$), and below the cardiology-resident tier ($0.862$), locating \sysname-2B within a clinically interpretable junior-reader range. Compared with existing ECG agents and zero-shot vision--language systems, this represents a substantial shift: the model is no longer only producing ECG-like text or isolated labels, but begins to align with the diagnostic behavior of human trainees on an external reader benchmark.

The per-condition results show where this alignment is strongest. \sysname-2B exceeded the medical-student tier on five of the six diagnoses---first-degree AV block ($0.818$ vs $0.732$), LBBB ($0.929$ vs $0.915$), sinus bradycardia ($0.778$ vs $0.750$), atrial fibrillation ($0.765$ vs $0.706$) and sinus tachycardia ($0.909$ vs $0.873$)---and exceeded the emergency-resident tier on three of the six; RBBB remains the single condition below both junior tiers ($0.841$ vs $0.928$ and $0.852$). These gains indicate that clinical diagnostic priors provide more than superficial report fluency: they help the agent recover clinically meaningful decision patterns that are comparable to early-stage human ECG readers.

\begin{finding}[title=Finding 3]
Clinical diagnostic priors move ECG agents toward human-reader performance. On CODE-test, \sysname-2B achieved a macro-F1 of $0.840$ and a micro-F1 of $0.852$, surpassing the medical-student ($0.817$) and emergency-resident ($0.829$) tiers and approaching the cardiology-resident tier ($0.862$), while exceeding medical-student performance on five of the six individual diagnoses.
\end{finding}

\subsection{Doctor-style supervision enables transferable ECG interpretation across cohorts}

Clinical ECG interpretation should remain stable beyond the dataset on which the agent is trained. A model that performs well only in-domain may have learned cohort-specific waveform shortcuts, device artifacts, or label conventions rather than transferable diagnostic structure. We therefore evaluated whether doctor-style supervision enables ECG interpretation to transfer across external cohorts with no training contact, spanning Brazil, China, and Germany.

Across the three external cohorts, \sysname-2B retained macro-F1 between $0.644$ and $0.840$ over the six CODE-6 red-flag diagnoses (cf. Fig.~\ref{fig:exp-human-transport}b). On CODE-test, the model reached macro-F1 $0.840$; on Chapman--Shaoxing, it retained $0.644$ despite the large cross-country shift; and on PTB-XL, it reached $0.678$. This stability was not limited to a single diagnosis. Atrial fibrillation remained strong across both Chapman and PTB-XL, with F1 $0.898$ and $0.892$, respectively, while sinus tachycardia also transferred consistently, with F1 $0.889$ on Chapman and $0.836$ on PTB-XL. Other general-purpose VLMs remained at the same low diagnostic level observed in the preceding report-generation and measurement evaluations, indicating that the cross-cohort gains are not obtained by generic visual-language capacity alone.

The advantage becomes clearer when compared with a waveform classifier trained on the same MIMIC source data. Although the ResNet1d-101 baseline is evaluated under the same external-cohort setting, its macro-F1 drops to $0.470$ on CODE-test, $0.350$ on Chapman--Shaoxing, and $0.049$ on PTB-XL. In contrast, \sysname-2B maintains substantially higher performance across all three cohorts, with gains of $+0.370$, $+0.294$, and $+0.629$ macro-F1, respectively. Thus, the central result is not merely improved accuracy on one benchmark, but stronger transport of ECG interpretation across cohort, geography, and label-distribution shift.

Together, these findings indicate that doctor-style supervision helps ECG agents move beyond dataset-specific recognition toward transferable clinical interpretation. By learning to organize ECG evidence through report-like diagnostic priors, \sysname-2B preserves clinically meaningful performance across external cohorts rather than relying only on source-dataset correlations.

\begin{figure}[!tbp]
    \centering
    \includegraphics[width=\linewidth]{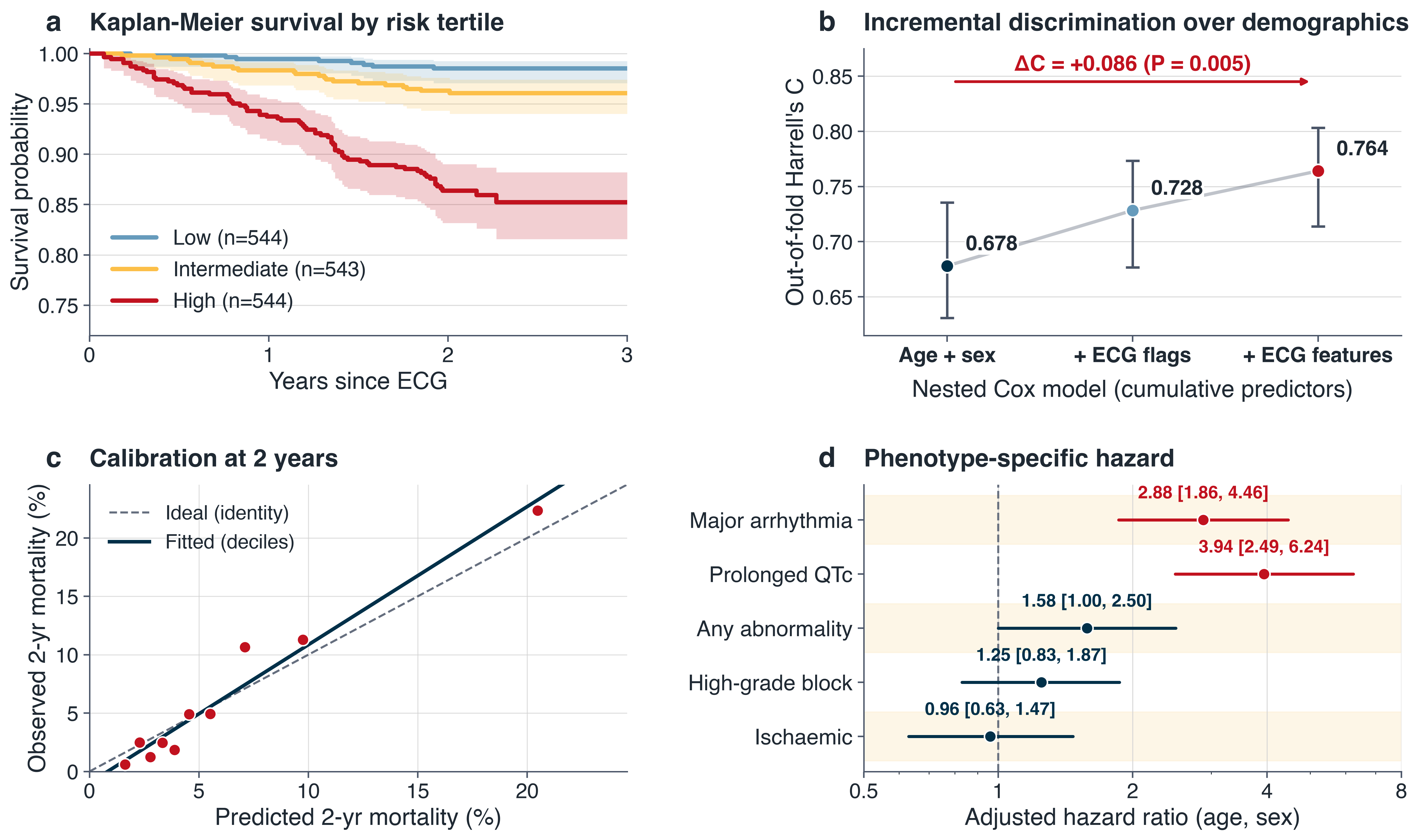}
    \caption{\footnotesize \textbf{Emergent prognostic value on the external Sami-Trop cohort
($n=1{,}631$; $104$ deaths; out-of-fold).}
\textbf{(a)} Kaplan--Meier survival across risk tertiles, with two-year mortality of $1.5\%$, $3.7\%$, and $13.0\%$ (log-rank $P<0.001$).
\textbf{(b)} Adding report-derived features increased Harrell's $C$ from $0.678$ for age and sex to $0.764$ ($\Delta C=+0.086$, $95\%$ CI $+0.026$ to $+0.147$; $P=0.005$).
\textbf{(c)} Two-year calibration across risk deciles (slope $1.18$, $r=0.98$).
\textbf{(d)} Age- and sex-adjusted hazard ratios for report-derived phenotypes, with significant associations for major arrhythmia and prolonged QTc.}
    \label{fig:prognosis}
\end{figure}

\begin{finding}[title=Finding 4]
Transferable ECG interpretation emerges from clinically structured supervision rather than source-dataset fitting alone. Across three external cohorts, \sysname-2B maintained macro-F1 scores between $0.644$ and $0.840$, while a waveform classifier trained on the same source data degraded sharply, reaching only $0.049$ on PTB-XL.
\end{finding}

\subsection{Structured ECG reports carry emergent prognostic information}

Clinical value ultimately depends on whether an ECG agent's outputs relate to patient outcomes, not only to reference labels. We therefore asked whether the structured reports produced by doctor-grounded supervision carry prognostic information beyond patient demographics, even though \sysname-2B was trained only to transcribe ECGs and never encountered mortality outcomes.

On the external Sami-Trop cohort (Chagas cardiomyopathy; $n=1{,}631$, $104$ deaths, median follow-up $2.1$\,years, no training contact), features extracted from \sysname-2B's reports added measurable prognostic value over age and sex. In an out-of-fold analysis, the concordance index of a demographic Cox model rose from $0.678$ to $0.764$ when the report features were added, an increment of $\Delta C = +0.086$ (paired bootstrap $95\%$ CI $+0.026$ to $+0.147$; $P=0.005$); the integrated discrimination improvement was $+0.047$ and the category-free net reclassification index $+0.36$, both excluding the null. This indicates that the diagnostic-transcription supervision encodes outcome-relevant structure recoverable in a cohort the model never saw.

The resulting risk score was well calibrated across risk deciles (calibration slope $1.18$; observed--predicted $r=0.98$) and stratified two-year mortality monotonically from $0.6\%$ in the lowest decile to $22.3\%$ in the highest (cf. Fig.~\ref{fig:prognosis}), with the signal concentrated in specific phenotypes---an age- and sex-adjusted hazard ratio of $2.88$ ($95\%$ CI $1.86$--$4.46$) for a major-arrhythmia flag and $3.94$ ($2.49$--$6.24$) for prolonged QTc---rather than an undifferentiated abnormal flag. This is an emergent, incremental association rather than a standalone risk predictor: demographics remain the dominant single covariate and adjustment is for age and sex only, so multivariable adjustment for comorbidities is required before clinical use. Together, these results show that doctor-grounded supervision produces reports whose structured content aligns with patient outcomes in a cohort with no training contact.

\begin{finding}[title=Finding 5]
Diagnostic supervision yields report representations that extend beyond label recovery. In a zero-contact cohort, report-derived features improved the age-and-sex Cox model from a C-index of $0.678$ to $0.764$, revealing emergent prognostic information even though \sysname-2B was never trained on mortality outcomes.
\end{finding}

\section{Discussion}
This study suggests that the path toward clinically useful ECG agents should not begin with larger architectures or more parameters, but with the way clinicians actually read ECGs. By embedding clinical reading priors into the supervision data, a compact 2B vision--language model can learn to generate report-like, measurement-grounded, human-reader-aligned, and transferable ECG interpretations without changing the underlying model architecture. The central implication is that supervision design, rather than model scale alone, can determine whether an ECG agent learns clinically meaningful interpretation or merely imitates visual-language patterns.

This distinction is especially important for medical agents. In clinical diagnosis, a prediction is rarely sufficient on its own; it must be supported by evidence that clinicians can inspect, question, and relate to established decision rules. For ECG interpretation, this means that an agent should not only output disease labels, but also organize findings in a doctor-like report, preserve core measurements, and connect waveform-derived evidence to diagnostic conclusions. Conventional classification supervision compresses this process into a final label, whereas doctor-style supervision exposes the intermediate structure of clinical reasoning. LuminaECG therefore treats ECG interpretation less as a generic recognition task and more as a structured reading task, where report language, quantitative measurements, and diagnostic decisions are learned together. This is the sense in which the model becomes more human-aligned: not because it reproduces human reasoning in full, but because its outputs are shaped around the same kinds of evidence that human readers use.

Placing the model on a human-reader scale further clarifies the meaning of its performance. Algorithm-only comparisons can show whether one system outperforms another, but they do not reveal whether the result is clinically interpretable. By comparing the agent with medical students, residents, and cardiologists on the same external benchmark, the evaluation provides a more meaningful reference point. The finding that clinical diagnostic priors bring the agent above the junior-reader tiers suggests that doctor-style supervision can narrow the gap between generic ECG agents and early-stage human readers. This should not be understood as replacing expert interpretation, but as evidence that a compact agent can begin to acquire clinically recognizable diagnostic behavior when trained from the perspective of how ECGs are actually read.

The transfer results extend this argument beyond a single benchmark. A common limitation of medical AI systems is that strong in-domain performance may reflect dataset-specific correlations, acquisition artifacts, or label conventions rather than stable clinical knowledge. ECG interpretation is particularly vulnerable to this problem because waveforms vary across devices, populations, institutions, and annotation systems. The improved cross-cohort behavior of LuminaECG suggests that doctor-style supervision encourages the model to learn more portable diagnostic structure. By grounding outputs in report-like evidence and clinical measurements, the agent is less dependent on source-dataset shortcuts and better able to preserve interpretation across external cohorts. This supports the broader view that human diagnostic priors can act as a form of transportable supervision.

Future work should strengthen the supervision beyond report-level imitation. Doctor-style reports provide an important clinical prior, but they do not fully expose the evidence and reasoning process behind each diagnosis. Incorporating lead-level evidence, morphology-specific descriptions, physician-verified rationales, and multimodal clinical context such as symptoms, medications, laboratory values, and prior ECGs would help the model move from reproducing clinical report structure toward evidence-grounded reasoning. The goal is not to make ECG agents memorize reporting templates, but to teach them how clinicians connect measurements, waveform morphology, lead distribution, and patient context to diagnostic decisions. This shift from report-like generation to explicit clinical reasoning is a key direction for building ECG agents that are more faithful, explainable, and useful in real-world settings.

Overall, LuminaECG points to a different design principle for medical vision--language agents. Rather than first asking how large the model should be, the more important question may be what kind of clinical reading process the model is asked to learn. For ECG interpretation, embedding human diagnostic priors into the data can produce an agent that writes more clinically meaningful reports, preserves quantitative evidence, approaches human trainee-level performance, and transfers better across cohorts. This suggests that the next stage of medical agent development should prioritize clinically structured supervision and human-readable evidence, not only architectural scale.

\section{Methods}
\label{sec:methods}

\subsection{Doctor-style ECG supervision construction}
\label{sec:supervision}

We build the training data from MIMIC-IV-ECG ($800{,}035$ twelve-lead recordings), split $9{:}1$ at the patient level into $719{,}828$ training and $79{,}981$ in-domain held-out records. The same construction is applied, with no retraining, to every external evaluation cohort (PTB-XL, Chapman--Shaoxing, CODE-test, Sami-Trop), so that cross-cohort differences reflect distribution shift rather than data preparation. Each record passes through three offline stages: measurement extraction, wave-annotated image rendering, and grounded report generation.

\paragraph{Measurement extraction.}
A NeuroKit-based delineator locates per-beat P, Q, R, S, T and U fiducial points across all twelve leads and computes a $303$-field per-record clinical feature panel ($194$ per-lead, $109$ global): per-lead wave amplitudes, ST level and slope, T-wave symmetry, intervals and QRS/T areas; heart rate and heart-rate-variability metrics; mean PR/QRS/QT intervals with prolongation flags; frontal-plane P, QRS and T axes; QTc by Bazett and Fridericia; regional ST-segment levels and elevation/depression counts; R-wave progression; and rule-based rhythm, bundle-branch, hypertrophy, ischaemia and pre-excitation scores, each with a confidence and its supporting leads. Every numerical quantity that the reference report later asserts about a record originates from this measurement step rather than from a free-text label.

\paragraph{Image rendering with wave-segment overlay.}
Each recording is rendered as a clinical-paper twelve-lead image on a calibration grid ($25$\,mm/s, $10$\,mm/mV) in the standard $6\times2$ layout, with a colour overlay marking the P (green), QRS (red) and T (blue) wave segments. The grid fixes the physical time/voltage scale in pixel space so that intervals and amplitudes become measurable distances; the overlay makes the P, QRS and T components locally identifiable, so the visual encoder receives a faithful clinical image together with a self-evident segmentation cue.

\paragraph{Grounded report generation.}
For each training record, gpt-4o receives the MIMIC machine measurements and the extracted numeric panel---not the image---and writes one eight-section Chinese report (rhythm \& rate; intervals; axis \& voltage; conduction \& arrhythmia; ST--T / ischaemia; signal quality; evidence points; conclusion). Because the generator never sees the waveform, a grounding protocol constrains it to assert only what the device diagnosis and the measurements determine: diagnoses derive from the device statements, with no diagnosis introduced that is absent from them and none contradicted; computed values serve as evidence attached to the diagnosis they support (first-degree AV block anchored to PR $>\!200$\,ms); only label-entailed morphology is asserted (an infarct wall with its fixed lead set and pathological Q waves); case-specific detail the label does not determine is omitted rather than fabricated; measurements a rhythm renders unreliable are flagged (the Bazett QTc under atrial fibrillation); and the evidence section links each conclusion to a measurement or a colour-identified wave. Over all $719{,}828$ training reports, every report contains the eight sections ($100.0\%$), a numeric interval ($99.8\%$), and a colour-identified wave in the evidence section ($98.9\%$). Native labels are harmonised to a common six-finding schema for cross-cohort evaluation (PTB-XL SCP codes, Chapman--Shaoxing SNOMED codes, and CODE-test adjudicated labels). Each training example is a triple (wave-annotated image, fixed eight-section instruction, grounded report); a low-rank adapter is the only trained component.

\subsection{LuminaECG model implementation}

LuminaECG was built on a Qwen2B vision--language backbone and fine-tuned using the doctor-style ECG supervision data constructed in the previous section. No task-specific architectural modules were added. This design was intentional: the study aimed to isolate the effect of clinical reading priors in the supervision data rather than attributing improvements to model scaling or architecture changes.

The model was fine-tuned for two epochs on eight NVIDIA H100 GPUs. Unless otherwise specified, preprocessing, optimization, and training hyperparameters followed the official Qwen training configuration. Each training instance consisted of an ECG-derived visual input and its paired structured clinical report, enabling the model to learn report-style ECG interpretation directly from the constructed supervision data.

\section{Conclusion}

We introduce \sysname, a compact ECG reasoning model that embeds doctor-grounded priors into the training data to preserve waveform anatomy, clinical measurements, and evidence-to-diagnosis relationships. Across held-out and external cohorts, \sysname improves report generation, quantitative measurement, critical-finding recovery, human-reader alignment, and cross-cohort generalization, while its structured reports also retain emergent prognostic information. These results suggest that reliable medical reasoning depends not only on model scale, but also on whether the data reflect how clinicians localize, measure, and interpret evidence. Although prospective validation, stronger faithfulness, and broader clinical evaluation remain necessary, \sysname supports a broader design principle: expert diagnostic priors should be treated as a foundation for medical multimodal models rather than expected to emerge from scale alone.

\bibliographystyle{plainnat}
\bibliography{ref}

\clearpage
\appendix
\renewcommand{\thetable}{A\arabic{table}}
\renewcommand{\thefigure}{A\arabic{figure}}
\setcounter{table}{0}
\setcounter{figure}{0}

\section{Complete Evaluation Results}
\label{app:full-eval}

Full numerical results are reported on the MIMIC-IV-ECG held-out test set ($n=79{,}981$). Table~\ref{tab:app-nlg-numeric} summarizes report-generation quality and numeric-measurement accuracy, Table~\ref{tab:app-vshuman} presents the CODE-test comparison with human readers across expertise tiers, and Table~\ref{tab:app-crossdata} reports cross-dataset generalization across three external cohorts spanning three continents. All systems, including \sysname-2B, pretrain-2B, Qwen3-VL-8B (zero-shot), ECG-R1-8B, GPT-5, GPT-5.4, Claude-4, Claude-Opus-4.7, Gemini-3.1-Pro, and Qwen3-VL-235B, are evaluated on the same complete $79{,}981$-record test set using the standard evaluation pipeline. All metrics are computed on the full-report text slice.

Table~\ref{tab:app-nlg-numeric} reports the metrics available for all systems. Several frontier baselines rarely produce parseable QT/QTc intervals or P/QRS/T electrical axes, preventing a consistent comparison on these fields. We therefore report these measurements only for \sysname-2B. The corresponding mean absolute errors are $7.82$~ms for QT, $8.91$~ms for QTc, $8.64^{\circ}$ for P-axis, $5.92^{\circ}$ for QRS-axis, and $9.97^{\circ}$ for T-axis.

\begin{table}[H]
  \centering
  \small
  \setlength{\tabcolsep}{6pt}
  \caption{Complete evaluation on the MIMIC-IV-ECG held-out test set
    ($n=79{,}981$). Top block: report-generation quality, character-level
    text similarity to the reference report (higher is better). Bottom block:
    numeric-measurement accuracy, mean absolute error of the values extracted
    from each report (lower is better; heart rate in bpm, intervals in ms).
    Bold marks the best value in each column.}
  \label{tab:app-nlg-numeric}

  {\centering\textbf{Report-generation quality (higher is better)}\par}
  \vspace{2pt}
  \begin{tabular}{l rrrr rr}
    \toprule
    Model & BLEU-1 & BLEU-2 & BLEU-3 & BLEU-4 & ROUGE-L & CIDEr \\
    \midrule
    pretrain-2B (ablation, no SFT) & 0.425 & 0.328 & 0.257 & 0.209 & 0.331 & 0.001 \\
    Qwen3-VL-8B (zero-shot)        & 0.262 & 0.221 & 0.184 & 0.155 & 0.393 & 0.000 \\
    ECG-R1-8B (external)           & 0.081 & 0.072 & 0.065 & 0.058 & 0.185 & 0.000 \\
    GPT-5.4                        & 0.594 & 0.466 & 0.374 & 0.307 & 0.392 & 0.029 \\
    GPT-5                          & 0.584 & 0.459 & 0.365 & 0.298 & 0.401 & 0.014 \\
    Claude-Opus-4.7                & 0.562 & 0.442 & 0.355 & 0.292 & 0.383 & 0.014 \\
    Claude-4                       & 0.379 & 0.317 & 0.265 & 0.224 & 0.419 & 0.000 \\
    Gemini-3.1-Pro                 & 0.000 & 0.000 & 0.000 & 0.000 & 0.067 & 0.000 \\
    Qwen3-VL-235B                  & 0.434 & 0.347 & 0.276 & 0.225 & 0.390 & 0.000 \\
    \midrule
    \textbf{\sysname-2B (ours)}    & \textbf{0.853} & \textbf{0.810} & \textbf{0.773} & \textbf{0.741} & \textbf{0.808} & \textbf{0.896}\\ 
    \bottomrule
  \end{tabular}

  \vspace{8pt}

  {\centering\textbf{Numeric-measurement accuracy (MAE, lower is better)}\par}
  \vspace{2pt}
  \begin{tabular}{l r r r}
    \toprule
    Model & HR (bpm) & PR (ms) & QRS (ms) \\
    \midrule
    
    pretrain-2B (ablation, no SFT) & 19.58 & 24.90 & 26.64 \\
    Qwen3-VL-8B (zero-shot)        & 15.73 & 44.99 & 16.09 \\
    ECG-R1-8B (external)           & 38.00 & 48.40 & 24.76 \\
    GPT-5.4                        & 6.45  & 24.79 & 20.24 \\
    GPT-5                          & 14.99 & 25.58 & 19.09 \\
    Claude-Opus-4.7                & 8.37  & 25.86 & 14.44 \\
    Claude-4                       & 17.08 & 24.97 & 21.05 \\
    Gemini-3.1-Pro                 & 8.87  & 23.97 & 15.20 \\
    Qwen3-VL-235B                  & 15.29 & 24.91 & 21.05 \\
    \midrule
    \textbf{\sysname-2B (ours)}    & \textbf{0.43}  & \textbf{8.97}  & \textbf{4.48} \\
    \bottomrule
  \end{tabular}
\end{table}

\begin{table}[H]
  \centering
  \small
  \setlength{\tabcolsep}{6pt}
  \caption{CODE-test ($n=827$) comparison of \sysname-2B against human readers of
    different expertise tiers and the Ribeiro-2020 deep-convolutional baseline,
    reported as per-condition F1 over the six CODE-6 red-flag diagnoses and as
    overall macro/micro-F1 (higher is better). Rows are ordered by macro-F1.
    \sysname-2B ranks above the medical-student and emergency-resident tiers
    overall and exceeds the medical-student tier on five of the six diagnoses.}
  \label{tab:app-vshuman}
  \begin{tabular}{l cccccc cc}
    \toprule
    Reader / Model & 1dAVb & RBBB & LBBB & SB & AF & ST & macro-F1 & micro-F1 \\
    \midrule
    Cardiologist~2        & 0.926 & 0.971 & 1.000 & 0.897 & 0.889 & 0.930 & 0.935 & 0.945 \\
    Ribeiro-DNN           & 0.897 & 0.944 & 1.000 & 0.882 & 0.870 & 0.960 & 0.925 & 0.938 \\
    Cardiologist~1        & 0.828 & 0.957 & 0.966 & 0.897 & 0.870 & 0.896 & 0.902 & 0.908 \\
    Cardiology resident   & 0.776 & 0.917 & 0.947 & 0.882 & 0.769 & 0.882 & 0.862 & 0.876 \\
    ED resident           & 0.719 & 0.852 & 0.912 & 0.848 & 0.696 & 0.946 & 0.829 & 0.846 \\
    Medical student       & 0.732 & 0.928 & 0.915 & 0.750 & 0.706 & 0.873 & 0.817 & 0.833 \\
    \midrule
    \textbf{\sysname-2B (ours)} & 0.818 & 0.841 & 0.929 & 0.778 & 0.765 & 0.909 & \textbf{0.840} & \textbf{0.852} \\
    \bottomrule
  \end{tabular}
\end{table}

\begin{table}[H]
  \centering
  \small
  \setlength{\tabcolsep}{5pt}
  \caption{Cross-dataset generalization on three external cohorts with no
    training contact, spanning three continents. Columns 2--7 give \sysname-2B
    per-condition F1 over the six CODE-6 conditions (SB, ST denote sinus
    bradycardia and sinus tachycardia); the last two columns give macro-F1 for
    \sysname-2B and for a plain ResNet1d-101 multi-label classifier trained on
    the same MIMIC waveforms (higher is better). \sysname-2B retains
    $0.64$--$0.84$ macro-F1 across continents, whereas the CNN, matched
    in-domain on MIMIC, collapses to $0.05$--$0.47$ out of distribution.}
  \label{tab:app-crossdata}
  \begin{tabular}{l cccccc c c}
    \toprule
     & \multicolumn{6}{c}{\sysname-2B per-condition F1} & \multicolumn{2}{c}{macro-F1} \\
    \cmidrule(lr){2-7} \cmidrule(lr){8-9}
    Cohort (region, $n$) & 1dAVb & RBBB & LBBB & SB & AF & ST & \sysname-2B & ResNet1d-101 \\
    \midrule
    CODE-test (Brazil, $827$)   & 0.818 & 0.841 & 0.929 & 0.778 & 0.765 & 0.909 & \textbf{0.840} & 0.470 \\
    Chapman (China, $45{,}151$) & 0.460 & 0.862 & 0.315 & 0.440 & 0.898 & 0.889 & \textbf{0.644} & 0.350 \\
    PTB-XL (Germany, $4{,}396$) & 0.430 & 0.638 & 0.760 & 0.512 & 0.892 & 0.836 & \textbf{0.678} & 0.049 \\
    \bottomrule
  \end{tabular}
\end{table}

\section{Case studies}
\label{app:cases}

\definecolor{ecgGreen}{HTML}{1C7A44}
\definecolor{ecgRed}{HTML}{B42318}
\definecolor{ecgNavy}{HTML}{2F6F9F}
\definecolor{ecgRefG}{HTML}{2F7D52}
\definecolor{ecgSlate}{HTML}{5A6774}
\newtcolorbox{sysbox}[2]{%
  enhanced, breakable=false, boxrule=0.7pt, arc=1.4mm,
  colback=white, colframe=#1,
  coltitle=white, colbacktitle=#1, fonttitle=\sffamily\bfseries\small,
  title={#2}, fontupper=\small,
  left=2.6mm, right=2.6mm, top=1.3mm, bottom=1.5mm, boxsep=1mm, before skip=5pt, after skip=5pt}
\newcommand{\ecgmiss}[1]{\textit{\textcolor{ecgSlate}{[missed: #1]}}}
\newtcolorbox{casebox}[1]{%
  enhanced, breakable, boxrule=0.9pt, arc=2mm,
  colback=white, colframe=black!38, coltitle=black!80, colbacktitle=black!8,
  fonttitle=\sffamily\bfseries, title={#1},
  left=2.8mm, right=2.8mm, top=1.8mm, bottom=2mm, before skip=9pt, after skip=9pt}
\newtcolorbox{runbox}[1]{%
  enhanced, breakable=false, boxrule=0.7pt, arc=1.4mm,
  colback=white, colframe=ecgNavy, coltitle=white, colbacktitle=ecgNavy,
  fonttitle=\sffamily\bfseries\small, title={#1},
  left=2.6mm, right=2.6mm, top=1.3mm, bottom=1.5mm, before skip=6pt, after skip=6pt}

\noindent We illustrate \sysname{} qualitatively in two parts. First, a head-to-head comparison
against four frontier zero-shot systems on two individual 12-lead ECGs. For each case we show the
tracing once, then every system's report---the physician reference, \sysname-2B, and the four
baselines (Qwen3-VL-235B, GPT-5.4, Claude-Opus-4.7, Gemini-3.1-Pro)---rendered verbatim in
translation and condensed to the diagnostically relevant sections. Within each report, correct
findings are shown in \textcolor{ecgGreen}{\textbf{green}}, incorrect or misattributed calls in
\textcolor{ecgRed}{\textbf{red}}, and findings present in the reference but missed are flagged in
\textcolor{ecgSlate}{grey}. Second, a set of stand-alone \sysname-2B readings of further held-out
ECGs, presented as an interactive session.

\medskip
\newpage
\noindent{\sffamily\bfseries B.1\quad Head-to-head against frontier systems}

\smallskip
\noindent{\footnotesize\textcolor{ecgGreen}{\rule{1.5ex}{1.5ex}}~correct\quad
\textcolor{ecgRed}{\rule{1.5ex}{1.5ex}}~incorrect / misattributed\quad
\textcolor{ecgSlate}{\textit{[missed]}}~present in the reference but absent from the read}

\smallskip
\begin{casebox}{Case A\hspace{0.7em}{\mdseries Atrial fibrillation, complete right bundle branch block, and prolonged QTc}}
{\footnotesize\textcolor{black!45}{MIMIC-IV-ECG, sample 15292609\_44181131}}

\begin{center}
  \includegraphics[width=0.9\linewidth]{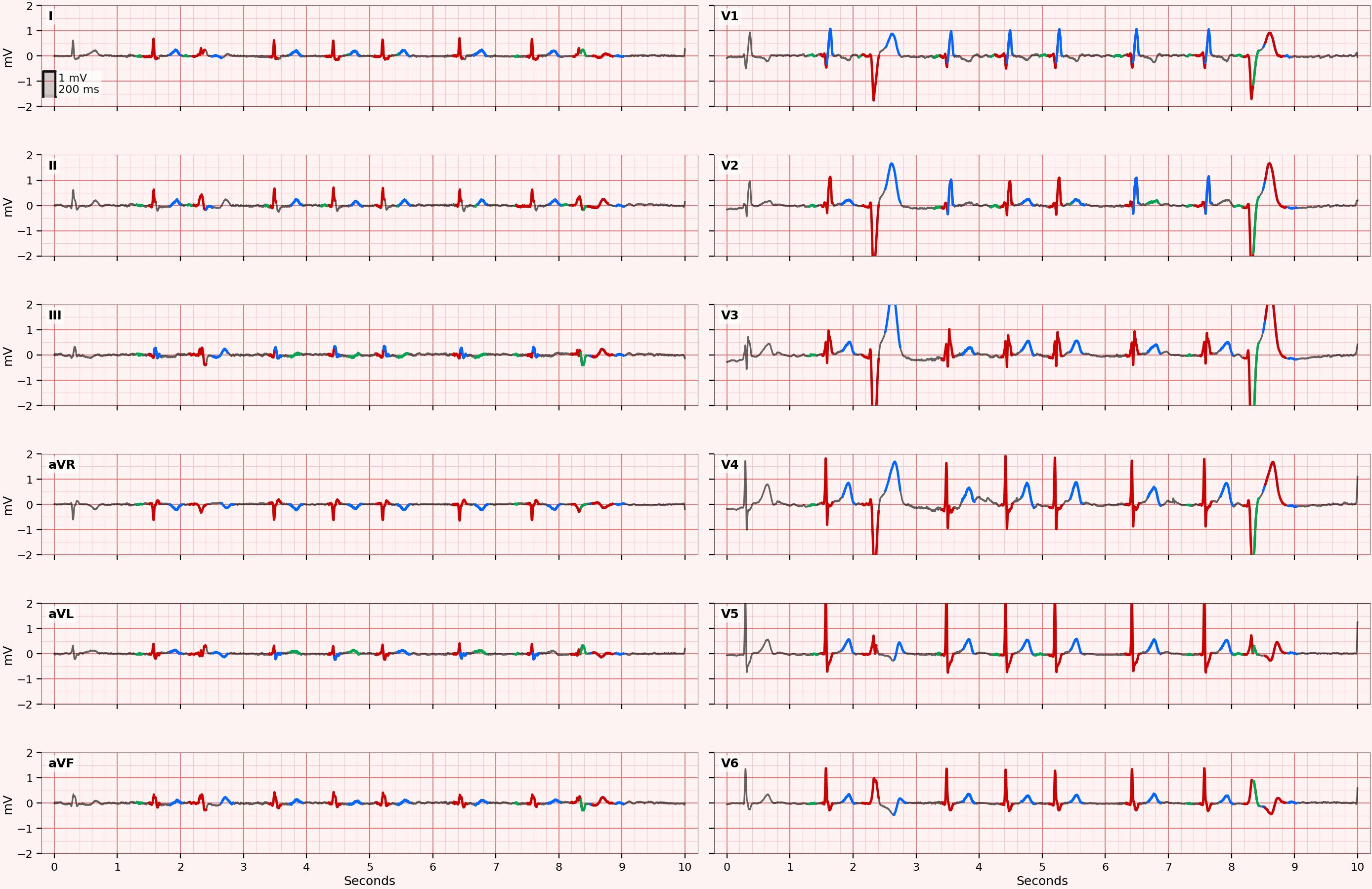}\\[0.2ex]
  {\footnotesize\textcolor{black!45}{Rendered 12-lead ECG with colour-coded P\,/\,QRS\,/\,T overlay.}}
\end{center}

\begin{sysbox}{ecgRefG}{Physician\hspace{0.5em}{\mdseries\small $\cdot$ reference $\cdot$ ground truth}}
\textbf{Rhythm.} \textcolor{ecgGreen}{Atrial fibrillation} with occasional PVCs / intraventricular conduction anomaly; ventricular rate $\sim$62\,bpm.\;
\textbf{Intervals.} PR not reliably assessable; QRS $\approx$144\,ms; QT $\approx$488\,ms; QTc $\approx$496\,ms (QRS markedly prolonged, QTc mildly prolonged).\;
\textbf{Conduction.} Consistent with \textcolor{ecgGreen}{complete right bundle branch block}; atrial fibrillation with occasional PVCs.\;
\textbf{ST-T.} No clear ischaemic or infarct-type ST-T change.\;
\textbf{Conclusion.} \textcolor{ecgGreen}{Atrial fibrillation}, \textcolor{ecgGreen}{complete right bundle branch block}, mildly prolonged QTc --- abnormal ECG.
\end{sysbox}

\begin{sysbox}{ecgNavy}{LuminaECG-2B\hspace{0.5em}{\mdseries\small $\cdot$ ours}}
\textbf{Rhythm.} \textcolor{ecgGreen}{Atrial fibrillation} with slow ventricular response; $\sim$59\,bpm.\;
\textbf{Intervals.} PR not reliably assessable; QRS $\approx$146\,ms; QT $\approx$490\,ms; QTc $\approx$486\,ms (QRS markedly prolonged, QTc near the upper limit).\;
\textbf{Conduction.} Consistent with \textcolor{ecgGreen}{complete right bundle branch block}; atrial fibrillation with slow ventricular response.\;
\textbf{Evidence.} RR intervals markedly irregular across leads with no clear green P waves $\rightarrow$ \textcolor{ecgGreen}{atrial fibrillation}; red QRS in V1 shows an rSR$'$ morphology with widening $\rightarrow$ \textcolor{ecgGreen}{right bundle branch block}; blue T waves mildly abnormal inferiorly, without acute-ischaemic ST change.\;
\textbf{Conclusion.} \textcolor{ecgGreen}{Atrial fibrillation} with slow ventricular response, \textcolor{ecgGreen}{complete right bundle branch block}, QTc near the upper limit --- abnormal ECG.
\end{sysbox}

\begin{sysbox}{ecgSlate}{Qwen3-VL-235B}
\textbf{Rhythm.} \textcolor{ecgRed}{Sinus rhythm}; $\sim$75\,bpm.\;
\textbf{Intervals.} PR $\approx$160\,ms; \textcolor{ecgRed}{QRS $\approx$80\,ms}; QT $\approx$380\,ms; QTc $\approx$410\,ms.\;
\textbf{Conduction.} \textcolor{ecgRed}{No AV block, bundle branch block, or ectopy; regular rhythm.}\;
\textbf{ST-T.} T-wave inversion with mild ST depression in V1--V3 $\rightarrow$ anteroseptal non-specific repolarization abnormality.\;
\textbf{Conclusion.} \textcolor{ecgRed}{Sinus rhythm}; anteroseptal repolarization abnormality. \ecgmiss{AF, complete RBBB}
\end{sysbox}

\begin{sysbox}{ecgSlate}{GPT-5.4}
\textbf{Rhythm.} \textcolor{ecgRed}{Sinus rhythm}; $\sim$52\,bpm.\;
\textbf{Intervals.} PR $\approx$160\,ms; \textcolor{ecgRed}{QRS $\approx$90\,ms}; QT $\approx$440\,ms; QTc $\approx$410\,ms.\;
\textbf{Conduction.} QRS not widened; no definite block or ectopy; overall \textcolor{ecgRed}{sinus bradycardia}.\;
\textbf{ST-T.} Mild J-point / ST elevation with tall symmetric T in V2--V5 $\rightarrow$ early-repolarization-like; no acute ischaemic change.\;
\textbf{Conclusion.} \textcolor{ecgRed}{Sinus bradycardia}; precordial early-repolarization-like ST-T change. \ecgmiss{AF, complete RBBB}
\end{sysbox}

\begin{sysbox}{ecgSlate}{Claude-Opus-4.7}
\textbf{Rhythm.} \textcolor{ecgRed}{Sinus rhythm} with early wide-complex beats (\textcolor{ecgRed}{PVCs}); $\sim$65\,bpm.\;
\textbf{Intervals.} PR $\approx$160\,ms; QRS $\approx$90\,ms (sinus), $\approx$160\,ms (ectopic); QT $\approx$380\,ms; QTc $\approx$395\,ms.\;
\textbf{Conduction.} \textcolor{ecgRed}{PVCs on a sinus background}; wide bizarre beats, no preceding P, morphology suggesting RVOT origin.\;
\textbf{ST-T.} ST at baseline with normal T in sinus beats; secondary post-PVC change only; no primary ischaemic change.\;
\textbf{Conclusion.} \textcolor{ecgRed}{Sinus rhythm with PVCs} (suggesting RVOT origin). \ecgmiss{AF, complete RBBB}
\end{sysbox}

\begin{sysbox}{ecgSlate}{Gemini-3.1-Pro}
\textbf{Rhythm.} \textcolor{ecgRed}{Sinus rhythm}; $\sim$60\,bpm (baseline).\;
\textbf{Intervals.} PR $\approx$160\,ms; QRS $\approx$90\,ms; QT $\approx$400\,ms; QTc $\approx$400\,ms.\;
\textbf{Conduction.} \textcolor{ecgRed}{Multifocal premature ventricular complexes.}\;
\textbf{ST-T.} No clear ischaemic change.\;
\textbf{Conclusion.} \textcolor{ecgRed}{Sinus rhythm with multifocal PVCs.} \ecgmiss{AF, complete RBBB}
\end{sysbox}
\end{casebox}

\clearpage
\begin{casebox}{Case B\hspace{0.7em}{\mdseries Old anteroseptal myocardial infarction on an otherwise normal sinus tracing}}
{\footnotesize\textcolor{black!45}{MIMIC-IV-ECG, sample 18490877\_49900084}}

\begin{center}
  \includegraphics[width=0.9\linewidth]{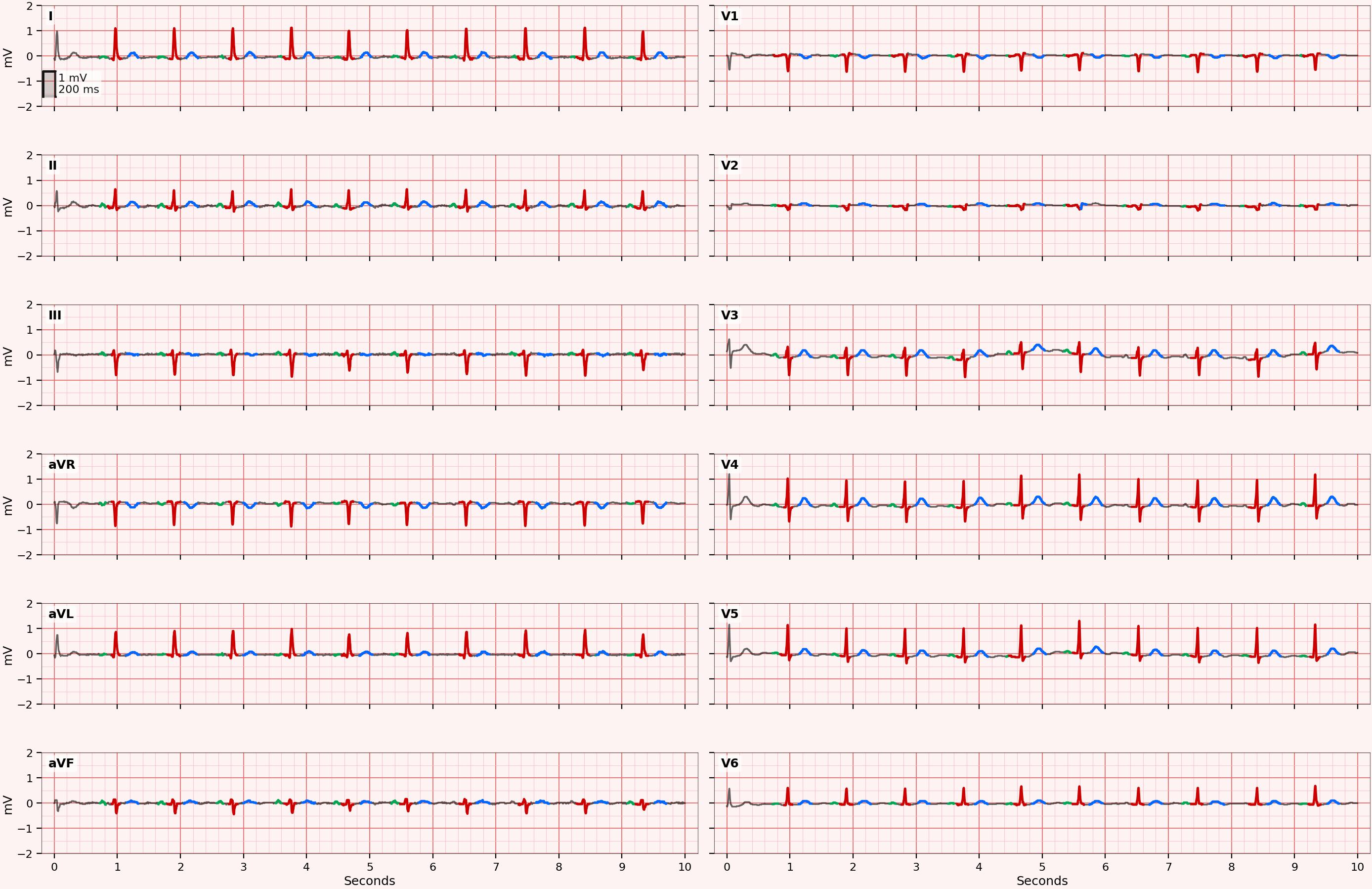}\\[0.2ex]
  {\footnotesize\textcolor{black!45}{Rendered 12-lead ECG with colour-coded P\,/\,QRS\,/\,T overlay.}}
\end{center}

\begin{sysbox}{ecgRefG}{Physician\hspace{0.5em}{\mdseries\small $\cdot$ reference $\cdot$ ground truth}}
\textbf{Rhythm.} Sinus rhythm; $\sim$65\,bpm, within normal range.\;
\textbf{Intervals.} PR $\approx$204\,ms; QRS $\approx$98\,ms; QT $\approx$434\,ms; QTc $\approx$452\,ms (all broadly within normal limits).\;
\textbf{Conduction.} AV and intraventricular conduction essentially normal; no definite block or significant arrhythmia.\;
\textbf{ST-T.} Suggests a possible \textcolor{ecgGreen}{old anteroseptal myocardial infarction}; correlate with clinical context and cardiac markers.\;
\textbf{Conclusion.} Sinus rhythm, normal rate and axis, no conduction block; suggests a possible \textcolor{ecgGreen}{old anteroseptal myocardial infarction} --- abnormal ECG.
\end{sysbox}

\begin{sysbox}{ecgNavy}{LuminaECG-2B\hspace{0.5em}{\mdseries\small $\cdot$ ours}}
\textbf{Rhythm.} Sinus rhythm; $\sim$65\,bpm, within normal range.\;
\textbf{Intervals.} PR $\approx$195\,ms; QRS $\approx$100\,ms; QT $\approx$431\,ms; QTc $\approx$449\,ms (broadly normal, QTc mildly long).\;
\textbf{Conduction.} AV and intraventricular conduction essentially normal; no definite block or significant arrhythmia.\;
\textbf{ST-T.} Suggests an \textcolor{ecgGreen}{old anteroseptal infarction}; no acute ischaemic ST-T change.\;
\textbf{Conclusion.} Sinus rhythm, normal rate and axis, no conduction block, QTc mildly long; suggests an \textcolor{ecgGreen}{old anteroseptal infarction} --- correlate with history and cardiac markers.
\end{sysbox}

\begin{sysbox}{ecgSlate}{Qwen3-VL-235B}
\textbf{Rhythm.} Sinus rhythm; $\sim$75\,bpm.\;
\textbf{Intervals.} PR $\approx$160\,ms; QRS $\approx$80\,ms; QT $\approx$380\,ms; QTc $\approx$410\,ms.\;
\textbf{Conduction.} No AV block, bundle branch block, or ectopy; regular rhythm.\;
\textbf{ST-T.} \textcolor{ecgRed}{No clear ischaemic change.}\;
\textbf{Conclusion.} \textcolor{ecgRed}{Normal ECG}; sinus rhythm, no conduction abnormality or ischaemia. \ecgmiss{old anteroseptal MI}
\end{sysbox}

\begin{sysbox}{ecgSlate}{GPT-5.4}
\textbf{Rhythm.} Sinus rhythm; $\sim$60\,bpm.\;
\textbf{Intervals.} PR $\approx$160\,ms; QRS $\approx$80\,ms; QT $\approx$400\,ms; QTc $\approx$400\,ms.\;
\textbf{Conduction.} AV and intraventricular conduction essentially normal; no block or significant arrhythmia.\;
\textbf{ST-T.} ST largely isoelectric; \textcolor{ecgGreen}{poor precordial R-wave progression, V1--V3 negative / low-voltage RS-QS}.\;
\textbf{Conclusion.} Sinus rhythm, left-axis deviation; \textcolor{ecgRed}{consider possible left anterior fascicular block}. \ecgmiss{old anteroseptal MI}
\end{sysbox}

\begin{sysbox}{ecgSlate}{Claude-Opus-4.7}
\textbf{Rhythm.} Sinus rhythm; $\sim$65\,bpm (RR $\approx$920\,ms, regular).\;
\textbf{Intervals.} PR $\approx$160\,ms; QRS $\approx$90\,ms; QT $\approx$400\,ms; QTc $\approx$418\,ms.\;
\textbf{Conduction.} No definite block, no ectopy; regular sinus rhythm.\;
\textbf{ST-T.} ST isoelectric, T concordant; \textcolor{ecgRed}{no ischaemic ST-T change and no pathological Q waves}.\;
\textbf{Conclusion.} \textcolor{ecgRed}{Normal sinus-rhythm ECG, no significant abnormality.} \ecgmiss{old anteroseptal MI}
\end{sysbox}

\begin{sysbox}{ecgSlate}{Gemini-3.1-Pro}
\textbf{Rhythm.} Sinus rhythm; $\sim$66\,bpm.\;
\textbf{Intervals.} PR $\approx$160\,ms; QRS $\approx$100\,ms; QT $\approx$400\,ms; QTc $\approx$420\,ms.\;
\textbf{Conduction.} No clear block or arrhythmia.\;
\textbf{ST-T.} No clear ischaemic change (shallow lead III T-inversion is a common variant).\;
\textbf{Conclusion.} \textcolor{ecgRed}{Sinus rhythm, largely normal ECG.} \ecgmiss{old anteroseptal MI}
\end{sysbox}
\end{casebox}

\medskip
\noindent{\sffamily\bfseries B.2\quad \sysname{} in action}

\smallskip
\noindent We show \sysname-2B reading four held-out 12-lead ECGs on its own, presented as an
interactive session: each tracing is submitted with a fixed instruction and the model returns its
report in eight structured steps; the readings are shown verbatim, translated from the model's
Chinese output.

\newcommand{\stepnum}[1]{\tikz[baseline=(sn.base)]{\node[circle,fill=ecgNavy,text=white,inner sep=0pt,minimum size=1.4em,font=\sffamily\bfseries\fontsize{7}{7}\selectfont](sn){#1};}}
\newcommand{\runex}[2]{%
\begin{minipage}[t]{0.325\linewidth}\vspace{0pt}\centering
  {\scriptsize\textcolor{black!50}{Rendered 12-lead ECG\\ P\,/\,QRS\,/\,T colour overlay}}\\[2.5pt]%
  \includegraphics[width=\linewidth]{#1}%
\end{minipage}\hfill
\begin{minipage}[t]{0.635\linewidth}\vspace{0pt}\footnotesize
  \setlength{\tabcolsep}{0pt}\renewcommand{\arraystretch}{1.32}%
  \begin{tabular}{@{}>{\centering\arraybackslash}p{1.6em}@{\hspace{1mm}}>{\bfseries\raggedright\arraybackslash}p{0.185\linewidth}>{\raggedright\arraybackslash}p{0.74\linewidth}@{}}#2\end{tabular}%
\end{minipage}}

\begin{runbox}{\sysname-2B\quad{\mdseries\small $\cdot$ held-out ECG, normal study}}
\runex{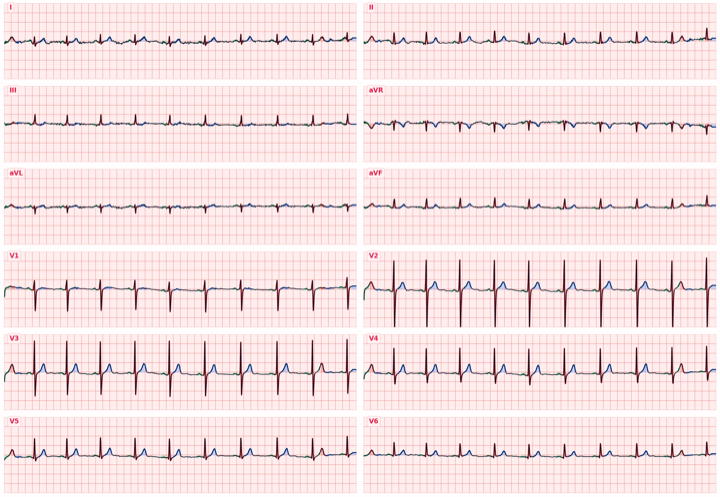}{%
\stepnum{1} & Rhythm & \textbf{Sinus rhythm}, $\sim$87\,bpm, mildly above the upper normal limit; regular RR intervals. \\
\stepnum{2} & Intervals & PR 156\,ms, QRS 90\,ms, QT 362\,ms, QTc 436\,ms (Bazett); all within usual normal ranges. \\
\stepnum{3} & Axis & P axis $+63^{\circ}$, QRS $+38^{\circ}$, T $+45^{\circ}$; normal frontal-plane axis; no voltage criteria for hypertrophy. \\
\stepnum{4} & Conduction & AV and intraventricular conduction within normal limits; no bundle-branch block, no significant arrhythmia. \\
\stepnum{5} & ST-T & No ischaemic or infarct-type ST-segment or T-wave change in any territory. \\
\stepnum{6} & Evidence & Lead II green P waves regularly precede the red QRS with consistent RR $\rightarrow$ sinus rhythm; blue T waves stable, no acute-ischaemic ST change. \\
\stepnum{7} & Signal & 0.005--150\,Hz diagnostic bandwidth with 60\,Hz notch; mild baseline drift and EMG noise; waveforms clear. \\
\stepnum{8} & Conclusion & \textbf{Sinus rhythm at a mildly increased rate, normal axis, no conduction block, no ischaemic ST-T change --- overall a normal ECG.} \\
}
\end{runbox}

\begin{runbox}{\sysname-2B\quad{\mdseries\small $\cdot$ held-out ECG, sinus tachycardia}}
\runex{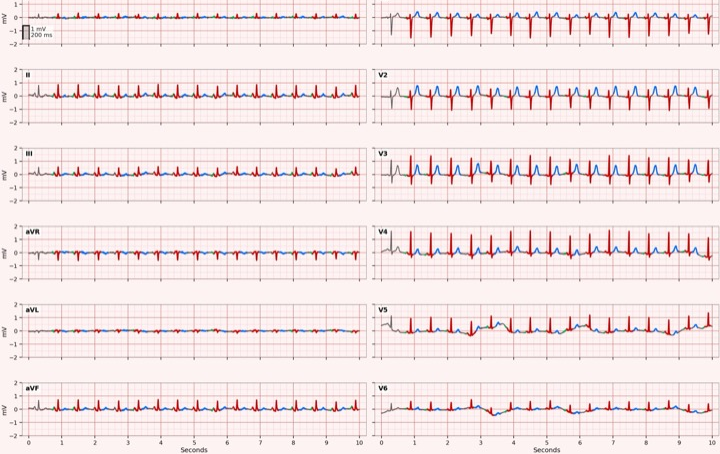}{%
\stepnum{1} & Rhythm & \textbf{Sinus tachycardia}, $\sim$100\,bpm; regular RR intervals with a consistent P--QRS relationship. \\
\stepnum{2} & Intervals & PR 122\,ms, QRS 80\,ms, QT 312\,ms, QTc 403\,ms; all within normal ranges. \\
\stepnum{3} & Axis & P axis $+78^{\circ}$, QRS $+71^{\circ}$, T $+77^{\circ}$; normal axis; no evidence of hypertrophy or voltage abnormality. \\
\stepnum{4} & Conduction & Normal AV and intraventricular conduction; no block and no significant arrhythmia beyond the tachycardia. \\
\stepnum{5} & ST-T & No ischaemic or infarct-type ST-T change. \\
\stepnum{6} & Evidence & Green P waves precede the red QRS with consistent RR $\rightarrow$ sinus origin of the tachycardia; blue T-wave morphology normal. \\
\stepnum{7} & Signal & Standard diagnostic bandwidth with 60\,Hz notch; low baseline drift and EMG noise; clear, interpretable waveforms. \\
\stepnum{8} & Conclusion & \textbf{Sinus tachycardia, normal axis, no conduction block, no ischaemic change --- overall a normal or near-normal ECG.} \\
}
\end{runbox}

\begin{runbox}{\sysname-2B\quad{\mdseries\small $\cdot$ held-out ECG, atrial fibrillation with rapid ventricular response}}
\runex{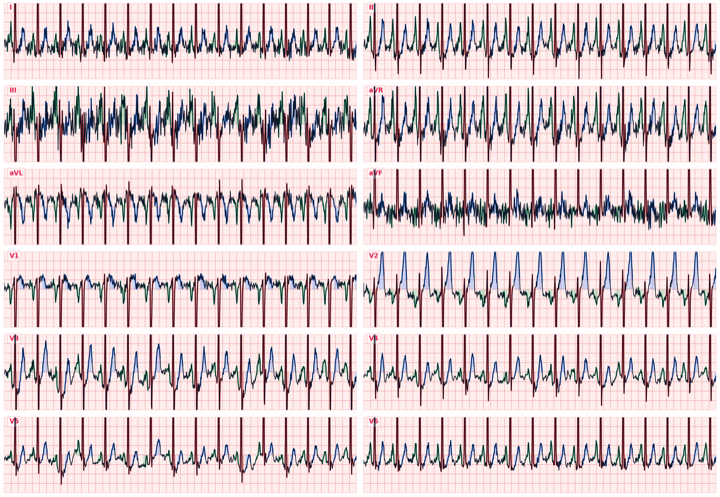}{%
\stepnum{1} & Rhythm & \textbf{Atrial fibrillation with rapid ventricular response}, $\sim$132\,bpm, clearly above the normal range. \\
\stepnum{2} & Intervals & PR not reliably assessable (no organised P waves); QRS 106\,ms; QT 310\,ms, \textbf{QTc 460\,ms (mildly prolonged)}. \\
\stepnum{3} & Axis & P axis not assessable; QRS $+66^{\circ}$, T $-166^{\circ}$; overall normal axis; no clear ventricular hypertrophy. \\
\stepnum{4} & Conduction & Consistent with \textbf{atrial fibrillation with rapid ventricular response}; no definite AV or intraventricular block. \\
\stepnum{5} & ST-T & \textbf{Diffuse ST-T changes suggesting possible myocardial ischaemia}; recommend correlation with clinical context and markers. \\
\stepnum{6} & Evidence & Markedly irregular RR across all leads, no clear green P waves $\rightarrow$ atrial fibrillation; blue T waves inverted or flattened in several leads $\rightarrow$ possible ischaemia. \\
\stepnum{7} & Signal & Waveforms clear with low baseline drift and EMG noise; adequate for rhythm and interval assessment. \\
\stepnum{8} & Conclusion & \textbf{Atrial fibrillation with rapid ventricular response, mildly prolonged QTc, and diffuse ST-T changes suggesting possible ischaemia --- abnormal ECG; recommend clinical correlation.} \\
}
\end{runbox}

\begin{runbox}{\sysname-2B\quad{\mdseries\small $\cdot$ held-out ECG, sinus bradycardia with LAE and LVH}}
\runex{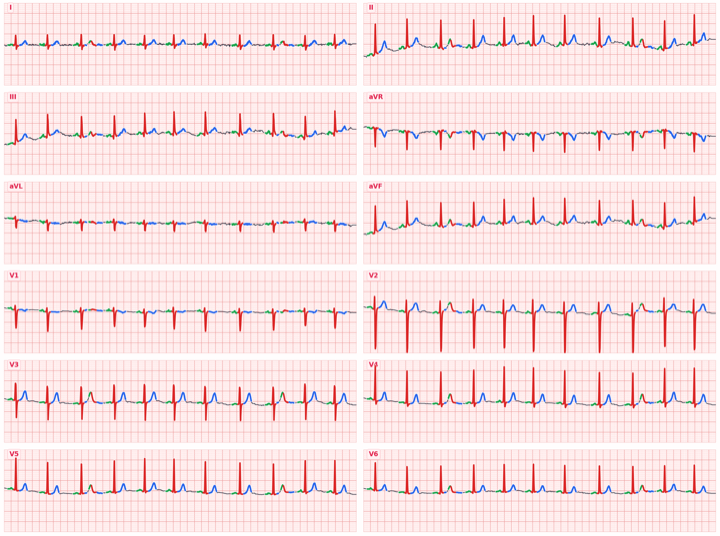}{%
\stepnum{1} & Rhythm & \textbf{Sinus bradycardia}, $\sim$56\,bpm, below the lower normal limit; regular RR intervals. \\
\stepnum{2} & Intervals & PR 187\,ms, QRS 110\,ms, QT 473\,ms, \textbf{QTc 457\,ms (mildly prolonged)}; PR and QRS roughly normal. \\
\stepnum{3} & Axis & P axis $+48^{\circ}$, QRS $+10^{\circ}$, T $+33^{\circ}$; normal axis; findings \textbf{suggest left atrial enlargement and left ventricular hypertrophy}. \\
\stepnum{4} & Conduction & AV and intraventricular conduction within normal limits; sinus bradycardia without significant arrhythmia. \\
\stepnum{5} & ST-T & \textbf{ST-segment elevation in some leads suggesting possible inferior injury}; anterior Q waves that may relate to LVH. \\
\stepnum{6} & Evidence & Lead II green P waves precede the red QRS with consistent RR $\rightarrow$ sinus rhythm; red QRS mildly widened $\rightarrow$ possible LVH; blue T with ST elevation $\rightarrow$ inferior-injury tendency. \\
\stepnum{7} & Signal & 0.0005--150\,Hz diagnostic bandwidth; signal acceptable, no obvious baseline drift or lead misplacement. \\
\stepnum{8} & Conclusion & \textbf{Sinus bradycardia with left atrial enlargement and LVH, mildly prolonged QTc, and possible inferior injury --- abnormal ECG; recommend clinical correlation.} \\
}
\end{runbox}

\end{document}